\documentclass[reprint,showpacs,preprintnumbers,amsmath,amssymb,prc,floatfix]{revtex4-1}
\usepackage{color}
\usepackage{graphicx}
\usepackage{dcolumn}
\usepackage{bm}
\usepackage[colorlinks,citecolor=blue,linkcolor=red,anchorcolor=blue,filecolor=blue,urlcolor=blue]{hyperref}

\begin{document}
\title{The surface properties of neutron-rich exotic nuclei within relativistic mean field formalisms}
\author{M. Bhuyan$^{1,2}$}
\email{Email: bhuyan@ita.br}
\author{B. V. Carlson$^1$}
\email{Email: brettvc@gmail.com}
\author{S. K. Patra$^3$}
\author{Shan-Gui Zhou$^{2,4,5}$}
\affiliation{$^1$Instituto Tecnol\'ogico de Aeron\'autica, 12.228-900 S\~ao Jos\'e dos Campos, S\~ao Paulo, 
Brazil}
\affiliation{$^2$Key Laboratory of Theoretical Physics, Institute of Theoretical Physics, Chinese Academy 
of Sciences, Beijing 100190, China} 
\affiliation{$^3$Institute of Physics, Sachivalaya Marg, Sainik School, Bhubaneswar 751005, India}
\affiliation{$^4$Center of Theoretical Nuclear Physics, National Laboratory of Heavy Ion Accelerator, Lanzhou 
730000, China}
\affiliation{$^5$Synergetic Innovation Center for Quantum Effects and Application, Hunan Normal University, 
Changsha, 410081, China}
\date{\today}

\begin{abstract}
In this theoretical study, we establish a correlation between the neutron skin thickness and the nuclear 
symmetry energy for the even$-$even isotopes of Fe, Ni, Zn, Ge, Se and Kr within the framework of the 
axially deformed self-consistent relativistic mean field for the non-linear NL3$^*$ and density-dependent 
DD-ME1 interactions. The coherent density functional method is used to formulate the symmetry energy, the 
neutron pressure and the curvature of finite nuclei as a function of the nuclear radius. We have performed 
broad studies for the mass dependence on the symmetry energy in terms of the neutron-proton asymmetry for 
mass 70  $\leq$ A $\leq$ 96. From this analysis, we found a notable signature of a shell closure at $N$ = 
50 in the isotopic chains of Fe, Ni, Zn, Ge, Se and Kr nuclei. The present study reveals a interrelationship 
between the characteristics of infinite nuclear matter and the neutron skin thickness of finite nuclei.
\end{abstract}


\maketitle

\section{Introduction}
Investigation of nuclei far from the line of beta stability has played a significant role in nuclear 
physics. Further, the advancement in the experimental facilities such as Jyav\~askyl\~a (Finland) 
\cite{leino95}, ORNL (United States) \cite{gross00}, CSR (China) \cite{sun03}, FAIR (Germany) 
\cite{wink08}, RIKEN (Japan) \cite{sakurai08}, GANIL (France) \cite{muller91}, GSI (Germany) 
\cite{geissel92}, FLNR (Russia) \cite{rodin03} and FRIB (United States) \cite{thoe10} has already opened 
new possibilities of exploring the production of various exotic nuclei and their properties under the 
extreme conditions of large isospin asymmetry. By virtue of the neutron-proton asymmetry in finite nuclei, 
one can gain insight into some of the basic components of the equation of state (EoS) of nuclear matter 
such as the symmetry energy $S_0$ ($\rho$) and the slope parameter $L_0$ ($\rho$) at nuclear saturation 
density $\rho_0$ \cite{tsang12}. In other words, the density properties of the symmetry energy of nuclear 
matter is forced to lie within a narrow window in terms of the nuclear bulk properties of neutron rich 
nuclei and vice-verse \cite{moll12}. In this context, a better understanding of the isospin and density 
dependent of the symmetry energy in exotic nuclei is one of the primary objectives of present studies.

Exploring the nuclear surface properties of semi-infinite nuclear matter is simple as compared to a 
finite nuclear system due to the absence of the many complexities arising from the shell, the charge, 
occupation probability and finite-size effects. In past works, the semi-classical Thomas-Fermi model 
\cite{myer85,kole85,cent98}, and quantal Hartree-Fock approach \cite{pear82,tond84} have investigated 
the surface properties of symmetric as well as asymmetric nuclear matter. Recently, considering the 
isotopic chains of doubly close shell nuclei, Warda {\it et al.} have demonstrated theoretically that 
the stiffness of the symmetry energy is dependent on the bulk and the surface component of the neutron 
skin thickness \cite{warda09,warda10,warda14}. Furthermore, the correlation between the volume and surface 
symmetry energy for finite nuclei has been extensively discussed in Refs. 
\cite{stein05,diep07,kolo08,kolo10,niko11,bks12}. In addition to these, the effects of temperature on the 
surface and the bulk symmetry have also been reported recently \cite{lee10,bks14}. In these works, the 
surface symmetry energy term is predicted to be more sensitive to the temperature than that of the volume 
one \cite{lee10,bks14}. The symmetry energy is not a directly measurable quantity but its value can be 
estimated indirectly from physical observables that are correlated to it. Two constraints are the 
experimental energy of the giant dipole resonance \cite{trip08} and the experimental differential 
cross-section data in a charge exchange reaction using the isospin dependent interaction of the optical 
potential \cite{khoa05,khoa07} (see the recent review of the Refs. \cite{she10,horo14} for details). 
Further, the connection of isospin asymmetry to the symmetry energy has an impact on many physical studies 
such as astrophysical observations \cite{li06,pika09,vida09,samm09,bhu13,bhu14}, the ground state structure 
of exotic nuclei \cite{niks08,giai10,van10,bhu15}, the determination of the neutron skin \cite{maza11,abra12}, 
the dynamics of heavy-ion reactions \cite{li08,chen08,colo09}, giant collective excitations 
\cite{rod07,subrat15}, the dipole polarizability \cite{rein10,tami11,piek12,roca13}, the mirror charge radii 
\cite{brow17,bao16}, the properties of compact star object \cite{stein05,bhu13,bhu14,pson07}, the 
nucleosynthesis process through neutrino convection \cite{bks12,robe12}, the photospheric radius of a neutron 
star \cite{bks12}, the core collapse of compact massive stars and related explosive phenomena at high density 
\cite{stein05,janka07}.

At present, a concerted effort has been put forth to determine the density properties of the symmetry 
energy and slope parameters for highly neutron-proton asymmetric systems in nuclear matter and in 
drip-line nuclei \cite{dan02,fami06,she07,cent09,gai07}. Broadly, these nuclear matter parameters are 
involved in the bulk properties of finite nuclei such as binding energies \cite{bks14,dan02,gai12}, 
relative nuclear radii \cite{bro00,typel01,gai11,gai12} and neutron density distributions 
\cite{dan02,fami06,gai11,gai12}. In this theoretical investigation, we study the relation between the 
neutron skin thickness and nuclear matter properties at saturation density, such as the symmetry energy, 
the neutron pressure and the curvature in an isotopic chain. Furthermore, we demonstrate in a few cases 
a relation between various bulk physical quantities of finite nuclei and the density properties 
of infinite nuclear matter. We consider the neutron rich even$-$even isotopes of medium mass nuclei such 
as Fe, Ni, Zn, Ge, Se and Kr in the present analysis, as, they are primary candidates in the the upcoming 
experimental facilities and several predictions have been made for them regarding the emergence of a 
nuclear skin. The calculations are performed within the axially deformed relativistic mean field approach, 
which has the ability to predict the nuclear skin thickness in exotic nuclei \cite{typel01,van92,typel14}. 
To interlink the infinite nuclear matter properties to intrinsic finite nuclear bulk properties, we have 
used the coherent density functional method \cite{gai07,gai12,anto79,anto80,anto94} through the energy 
density functional of Brueckner {\it et al.} \cite{brue68,brue69}. Briefly, our aim to constrain the 
nuclear matter observables using the inherent properties of exotic neutron rich finite nuclei as well as 
the contrary.

This paper is organized as follows: In Section II we discuss the theoretical model for the relativistic 
mean field approach along with coherent density functional method. Section III is assigned to the discussion 
of the results obtained from our calculation and of the possible correlation among the infinite nuclear matter 
and finite nucleus properties. Finally, a summary and a brief conclusion are given in Section IV.

\section{Theoretical formalism}
In the present work, we estimate the nuclear symmetry energy $S_0 (\rho)$, neutron pressure $p_0 (\rho)$ and 
other related physical quantities of exotic finite nuclei as functions of the baryon density on the basis of 
the corresponding definitions for asymmetric nuclear matter. We have taken a general form of the non-linear 
finite-range relativistic mean field model, considering it to be represented by the Lagrangian density (given 
in the next subsection) \cite{bogu77,sero86,typel01}. This model has been widely used to describe infinite 
nuclear matter, finite nuclei, and stellar matter properties for extreme isospin asymmetry 
\cite{bogu77,sero86,ring86,lala99c,bhu09,bhu11,rein89,ring96,vret05,meng06,paar07,niks11,logo12,zhao12,typel01,fuch95,niks02,type99,lala05,hoff01,bret00}. 
To calculate the effective intrinsic nuclear matter quantities in finite nuclei, one must know the key 
parameters of nuclear matter that characterize its density dependence at saturation density. The most general 
form of the nuclear matter symmetry energy $S (\rho)$ for the relativistic mean field models can be expressed 
as, 
\begin{eqnarray}
S^{NM} (\rho) = \frac{1}{8} \left(\frac{\partial^2 (\cal{E}/ \rho)}{\partial y^2} \right)_{\rho, y=1/2},
\label{sym}
\end{eqnarray}
where $y$ is the proton fraction for asymmetric nuclear matter. Here the detailed calculations of the 
energy density $\cal{E}$ as a function of density from the relativistic Lagrangian are given in Refs. 
\cite{bhu13,dutra14,dutra15,type99,niks02}. The widely used slope parameter $L^{NM}$ at saturation 
density is given as,
\begin{eqnarray}
L_0^{NM}= 3\rho\left(\frac{\partial S^{NM}}{\partial \rho} \right)_{\rho=\rho_0} = \frac{3p_0^{NM}}{\rho_0},
\label{slope}
\end{eqnarray}
where, $p_0^{NM}$ is the neutron pressure of nuclear matter at saturation density, $\rho_0$ being the 
saturation density of the symmetric nuclear matter.  Further, the curvature and skewness of the symmetry 
energy are given by,
\begin{eqnarray}
K_0^{NM}=9\rho^2\left(\frac{\partial^2 S^{NM}}{\partial\rho^2} \right)_{\rho=\rho_0},
\label{comp}
\end{eqnarray}
and
\begin{eqnarray}
Q_0^{NM}=27\rho^3\left(\frac{\partial^3 S^{NM}}{\partial \rho^3}\right)_{\rho=\rho_0},
\label{skew}
\end{eqnarray}
respectively. Our present knowledge of the basic properties of the symmetry energy around saturation density 
is still poor with its value estimated to be  about 27$\pm3$ MeV \cite{dutra12,dutra14}. In practice, this 
ambiguity play an essential role in the structure calculations of finite nuclei. Here, to obtain a general 
idea of what one might expect, we have used the calculated saturation properties of infinite nuclear matter 
from the relativistic mean field with non-linear NL3$^*$ and density-dependent DD-ME1 interaction parameters, 
which are listed in Table \ref{tab1} (for details see the Refs. \cite{bhu13,lala09,type99,niks02}). In the 
relativistic mean field (RMF) model, there is a strong correlation between the Dirac effective nucleon mass 
at saturation density and the strength of the spin-orbit force in finite nuclei \cite{sero97,ring90}. Further, 
one of the most compelling features of RMF models is the reproduction of the spin-orbit splittings in finite 
nuclei. This occurs when the velocity dependence of the equivalent central potential that leads to saturation 
arises primarily due to a reduced nucleon effective mass \cite{furn98}. On the other hand, the non-relativistic 
effective mass parametrizes the momentum dependence of the single-particle potential, which is the result of 
a quadratic parametrization of the single-particle spectrum.  It has been argued \cite{jami89} that the 
so-called Lorentz mass should be compared with the non-relativistic effective mass extracted from analyses 
carried out in the framework of non-relativistic optical and shell models.  

\subsection{The relativistic mean-field theory}
The fundamental theory of the strong interaction that can provide a complete description of nuclear equation 
of state is quantum chromodyanmics (QCD). At present, it is not conceivable to describe the complete picture 
of hadronic matter due to its non-perturbative properties. Hence, one needs to apply the perspective of an 
effective field theory (EFT) at low energy, such as quantum hadrodynamics (QHD) \cite{bogu77,sero86,ring86}. 
The mean field treatment of QHD has been used widely to describe the properties of infinite nuclear matter 
\cite{sero86,bhu13,bhu14,type99} and finite nuclei \cite{bogu77,ring86,bhu09,bhu11,bhu15,niks02,bret00}. 
In the relativistic mean field approach, the nucleus is considered as a composite system of nucleons (proton 
and neutron) interacting through the exchange of mesons and photons 
\cite{bend03,sero86,rein89,ring96a,vret05,meng06,paar07}. Further, the contributions from the meson fields 
are described either by mean fields or by point-like interactions between the nucleons \cite{niko92,bur02}. 
Density dependent coupling constants \cite{fuch95,niks02,type99,lala05,hoff01,bret00} and/or nonlinear 
coupling terms \cite{bogu77,bro92} are introduced to reproduced the properties of finite nuclei and the 
correct saturation properties of infinite nuclear matter. Here, most of the computational effort is devoted 
to solving the Dirac equation and calculating various densities. In the present calculation, we have used 
the microscopic self-consistent relativistic mean field (RMF) theory as a standard tool to investigate 
nuclear structure. It is worth mentioning that the RMF approach is one of the most popular and widely used 
formalisms. A typical relativistic Lagrangian density (after several modifications of the original Walecka 
Lagrangian to take care of various limitations) for a nucleon-meson many body system has the form 
\cite{bogu77,sero86,lala99c,bhu09,bhu11,rein89,ring96,vret05,meng06,paar07,niks11,logo12,zhao12},
\begin{eqnarray}
{\cal L}&=&\overline{\psi}\{i\gamma^{\mu}\partial_{\mu}-M\}\psi +{\frac12}\partial^{\mu}\sigma
\partial_{\mu}\sigma \nonumber \\ 
&& -{\frac12}m_{\sigma}^{2}\sigma^{2}-{\frac13}g_{2}\sigma^{3} -{\frac14}g_{3}\sigma^{4}
-g_{s}\overline{\psi}\psi\sigma \nonumber \\ 
&& -{\frac14}\Omega^{\mu\nu}\Omega_{\mu\nu}+{\frac12}m_{w}^{2}\omega^{\mu}\omega_{\mu}
-g_{w}\overline\psi\gamma^{\mu}\psi\omega_{\mu} \nonumber \\
&&-{\frac14}\vec{B}^{\mu\nu}.\vec{B}_{\mu\nu}+\frac{1}{2}m_{\rho}^2\vec{\rho}^{\mu}.\vec{\rho}_{\mu} 
-g_{\rho}\overline{\psi}\gamma^{\mu}\vec{\tau}\psi\cdot\vec{\rho}^{\mu} \nonumber \\
&&-{\frac14}F^{\mu\nu}F_{\mu\nu}-e\overline{\psi} \gamma^{\mu}\frac{\left(1-\tau_{3}\right)}{2}\psi A_{\mu}, 
\label{lag}
\end{eqnarray}
with vector field tensors
\begin{eqnarray}
F^{\mu\nu} = \partial_{\mu} A_{\nu} - \partial_{\nu} A_{\mu} \nonumber \\
\Omega_{\mu\nu} = \partial_{\mu} \omega_{\nu} - \partial_{\nu} \omega_{\mu} \nonumber \\
\vec{B}^{\mu\nu} = \partial_{\mu} \vec{\rho}_{\nu} - \partial_{\nu} \vec{\rho}_{\mu}.
\end{eqnarray}
Here the field for the $\sigma$-meson is denoted by $\sigma$, that for the $\omega$-meson by $\omega_{\mu}$, 
and for the isovector $\rho$-meson by $\vec{\rho}_{\mu}$. The electromagnetic field is defined by $A_{\mu}$.
The quantities, $\Omega^{\mu\nu}$, $\vec{B}_{\mu\nu}$, and $F^{\mu\nu}$ are the field tensors for the 
$\omega^{\mu}$, $\vec{\rho}_{\mu}$ and photon fields, respectively. 

The RMF model proposed in Refs. \cite{type99,niks02} allows density dependence of the meson-nucleon coupling, 
which is parametrized in a phenomenological approach \cite{fuch95,niks02,type99,lala05,hoff01,bret00}. The 
coupling of the mesons to the nucleon fields are defined as 
\begin{eqnarray}
g_i (\rho) = g_i (\rho_{sat}) f_i (x) \vert_{i = \sigma, \omega},  
\end{eqnarray}
where, 
\begin{eqnarray}
f_i (x) = a_i \frac{1+b_i(x + d_i)^2}{1+c_i(x+d_i)^2},
\label{funf}
\end{eqnarray}
and 
\begin{eqnarray}
g_{\rho} = g_{\rho} (\rho_{sat})e^{a_{\rho}(x-1)}.
\end{eqnarray}
Here, the functional $x = \rho/\rho_{sat}$ and the eight real parameters in Eq. (\ref{funf}) are not 
independent. The five constraints $f_i (1) = 1$, $f''_{\sigma} (1) = f''_{\omega}$ (1) and $f''_i (0)= 0$ 
reduce the number of independent parameters to three. These independent parameters (coupling parameters 
and the mass of the $\sigma$ meson) were adjusted to reproduce the properties of symmetric and asymmetric 
nuclear matter and the ground state properties of finite nuclei.

From the above Lagrangian density we obtain the field equations for the nucleons and the mesons. These 
equations are solved by expanding the upper and lower components of the Dirac spinors and the boson fields 
in an axially deformed harmonic oscillator basis, with an initial deformation $\beta_{0}$. The set of 
coupled equations is solved numerically by a self-consistent iteration method. The center-of-mass motion 
energy correction is estimated by the usual harmonic oscillator formula $E_{c.m.}=\frac{3}{4}(41A^{-1/3})$. 
The quadrupole deformation parameter $\beta_2$ is evaluated from the resulting proton and neutron quadrupole 
moments, as
\begin{equation}
Q=Q_n+Q_p=\sqrt{\frac{16\pi}5} (\frac3{4\pi} AR^2\beta_2).
\end{equation}
The root mean square (rms) matter radius is defined as
\begin{equation}
\langle r_m^2\rangle={1\over{A}}\int\rho(r_{\perp},z) r^2d\tau,
\end{equation}
where $A$ is the mass number, and $\rho(r_{\perp},z)$ is the deformed density. The total binding energy and 
other observables are also obtained by using the standard relations, given in Ref. \cite{ring90}. Here, we 
have used the NL3$^*$ \cite{lala09,lala97} and the density-dependent DD-ME1 \cite{niks02} interactions. 
These interactions are able to reproduce reasonably well the properties of not only the stable nuclei but 
also those not too far from the $\beta$-stability valley \cite{lala97,lala09,niks02,bret00}. In the outputs, 
we obtain the potentials, densities, single-particle energy levels, nuclear radii, deformations and the 
binding energies. For a given nucleus, the maximum binding energy corresponds to the ground state and other 
solutions are obtained as various excited intrinsic states at other deformations, provided the nucleus does 
not undergo fission. 

To describe the nuclear bulk properties of open-shell nuclei, one has to consider the pairing correlations 
in their ground as well as excited states \cite{karat10}. There are various methods, such as the BCS approach, 
the Bogoliubov transformation and particle number conserving methods, that have been developed to treat 
pairing effects in the study of nuclear properties including fission barriers 
\cite{zeng83,moli97,zhang11,hao12}. In principle, the Bogoliubov transformation is the most widely used 
method to take pairing correlations into account for the drip-line region 
\cite{vret05,paar07,ring96a,meng06,lala99a,lala99b}. In the case of nuclei not too far from the 
$\beta$-stability line, one can use the constant gap BCS pairing approach to obtain a reasonably good 
approximation of pairing \cite{doba84}. In the  present analysis, we have employed the constant gap BCS 
approach with the NL3$^*$ and a Bogoliubov transformation with DD-ME1 interactions 
\cite{mad81,moll88,bhu09,bhu15,niks02,typel01,bret00}.
\begin{table}
\caption{Parameters and infinite nuclear matter properties at saturation density of the non-linear NL3$^*$ 
\cite{lala09} and density-dependent DD-ME1 \cite{niks02} interaction parameters.}
\renewcommand{\tabcolsep}{0.6cm}
\renewcommand{\arraystretch}{1.2}
\begin{tabular}{llllllllll}
\hline \hline
NL3$^*$ interaction \cite{lala09} & DD-ME1 interaction \cite{niks02} \\
\hline
$M$ = 939               & $M$ = 939 \\
$m_{\sigma}$ = 502.5742 & $m_{\sigma}$ = 549.5255 \\
$m_{\omega}$ = 782.6000 & $m_{\omega}$ = 783.0000 \\
$m_{\rho}$ = 763.000    & $m_{\rho}$ = 763.000 \\
$m_{\sigma}$ = 10.0944  & $m_{\sigma}$ ($\rho_{sat}$) = 10.4434 \\
$m_{\omega}$ = 12.8065  & $m_{\omega}$ ($\rho_{sat}$) = 12.8939 \\
$m_{\rho}$ = 4.5748     & $m_{\rho}$ ($\rho_{sat}$) = 3.8053 \\
$g_2$ = -10.8093        & $a_{\sigma}$ = 1.3854 \\
$g_3$ = -30.1486        & $b_{\sigma}$ = 0.9781 \\
$M/M^*$ = 0.594         & $c_{\sigma}$ = 1.5342 \\ 
$\rho_0$ = 0.150        & $d_{\sigma}$ = 0.4661 \\
$\cal{E/A}$ = -16.31    & $a_{\omega}$ = 1.3879 \\
$K_0^{NM}$ = 258.27     & $b_{\omega}$ = 0.8525 \\
$S^{NM}$  = 38.68       & $c_{\omega}$ = 1.3566 \\
                        & $d_{\omega}$ = 0.4957 \\ 
                        & $a_{\rho}$ = 0.5008 \\
                        & $M/M^*$ = 0.586 \\
                        & $\rho_0$ = 0.152 \\
                        & $\cal{E/A}$ = -16.04 \\
                        & $K_0^{NM}$ = 244.72 \\
                        & $S^{NM}$  = 33.06 \\
\hline\hline      
\end{tabular}
\label{tab1}
\end{table}
\begin{table*}
\caption{The binding energy (BE), charge radius $r_{ch}$ and the quadrupole deformation parameter $\beta_2$ 
for the ground states of the $^{72-86}$Fe, $^{74-88}$Ni and $^{76-90}$Zn nuclei from the non-linear NL3$^*$ 
and the density dependent DD-ME1 calculations compare with the experimental data \cite{audi12,ange13,prit12}, 
wherever available. The (*) marks in the binding energies of the experimental data are for extrapolated 
values.}
\renewcommand{\tabcolsep}{0.12cm}
\renewcommand{\arraystretch}{1.4}
\begin{tabular}{cccccccccc}
\hline\hline
Nucleus & \multicolumn{3}{c}{Binding Energy} & \multicolumn{3}{c}{Charge Radius} 
& \multicolumn{3}{c}{Quadrupole Deformation} \\
& NL3$^*$ & DD-ME1 & Expt. \cite{audi12} & NL3$^*$ & DD-ME1 & Expt. \cite{ange13} 
& NL3$^*$ & DD-ME1 & Expt. \cite{prit12} \\
\hline \hline
$^{70}$Fe &580.68&580.59&577.43$^*$&3.875&3.879& $--$ &0.179&0.190& $--$ \\
$^{72}$Fe &588.89&588.68&589.10$^*$&3.898&3.899& $--$ &0.207&0.214& $--$ \\
$^{74}$Fe &594.98&594.78& $--$     &3.912&3.909& $--$ &0.198&0.188& $--$ \\
$^{76}$Fe &600.61&600.58& $--$     &3.938&3.935& $--$ &0.004&0.002& $--$ \\
$^{78}$Fe &602.99&603.65& $--$     &3.956&3.950& $--$ &0.261&0.254& $--$ \\
$^{80}$Fe &605.64&606.70& $--$     &3.958&3.966& $--$ &0.211&0.248& $--$ \\
$^{82}$Fe &608.40&609.21& $--$     &3.978&3.979& $--$ &0.223&0.225& $--$ \\
$^{84}$Fe &609.37&611.02& $--$     &3.994&3.992& $--$ &0.202&0.188& $--$ \\
$^{86}$Fe &610.40&612.64& $--$     &4.006&3.995& $--$ &0.174&0.124& $--$ \\
\hline
\hline
$^{72}$Ni &611.78&612.34&613.15     &3.901&3.892&$--$ &0.042&0.014& $--$ \\
$^{74}$Ni &621.94&622.31&623.74$^*$ &3.923&3.908&$--$ &0.099&0.096& 0.21$^*$ \\
$^{76}$Ni &630.95&631.52&633.16$^*$ &3.929&3.923&$--$ &0.009&0.006& $--$ \\
$^{78}$Ni &635.85&635.96&641.94$^*$ &3.945&3.935&$--$ &0.001&0.002& $--$ \\
$^{80}$Ni &643.51&643.61& $--$      &3.958&3.954&$--$ &0.008&0.011& $--$ \\
$^{82}$Ni &646.72&645.74& $--$      &3.974&3.967&$--$ &0.091&0.096& $--$ \\
$^{84}$Ni &649.67&650.64& $--$      &3.990&3.982&$--$ &0.085&0.059& $--$ \\
$^{86}$Ni &652.52&653.84& $--$      &3.995&3.994&$--$ &0.067&0.035& $--$ \\
$^{88}$Ni &655.08&656.76& $--$      &4.007&4.010&$--$ &0.034&0.005& $--$ \\
\hline
\hline
$^{74}$Zn &637.64&637.27&639.51     &3.985&3.981&$--$ &0.161&0.185& $--$ \\
$^{76}$Zn &650.78&650.51&652.08     &4.001&3.997&$--$ &0.182&0.201& $--$ \\
$^{78}$Zn &661.39&661.01&663.44     &4.008&4.006&$--$ &0.150&0.164& $--$ \\
$^{80}$Zn &670.90&670.99&674.08     &4.009&4.010&$--$ &0.001&0.002& $--$ \\
$^{82}$Zn &677.13&677.08&680.84$^*$ &4.039&4.042&$--$ &0.151&0.186& $--$ \\
$^{84}$Zn &682.51&682.59& $--$      &4.069&4.069&$--$ &0.202&0.216& $--$ \\
$^{86}$Zn &687.54&687.68& $--$      &4.096&4.095&$--$ &0.230&0.238& $--$ \\
$^{88}$Zn &691.09&691.76& $--$      &4.119&4.117&$--$ &0.228&0.227& $--$ \\
$^{90}$Zn &694.32&695.42& $--$      &4.136&4.136&$--$ &0.213&0.206& $--$ \\
\hline \hline
\end{tabular}
\label{Tab2}
\end{table*}

\subsection{The coherent density functional method}
\begin{figure}
\begin{center}
\includegraphics[width=1.00\columnwidth]{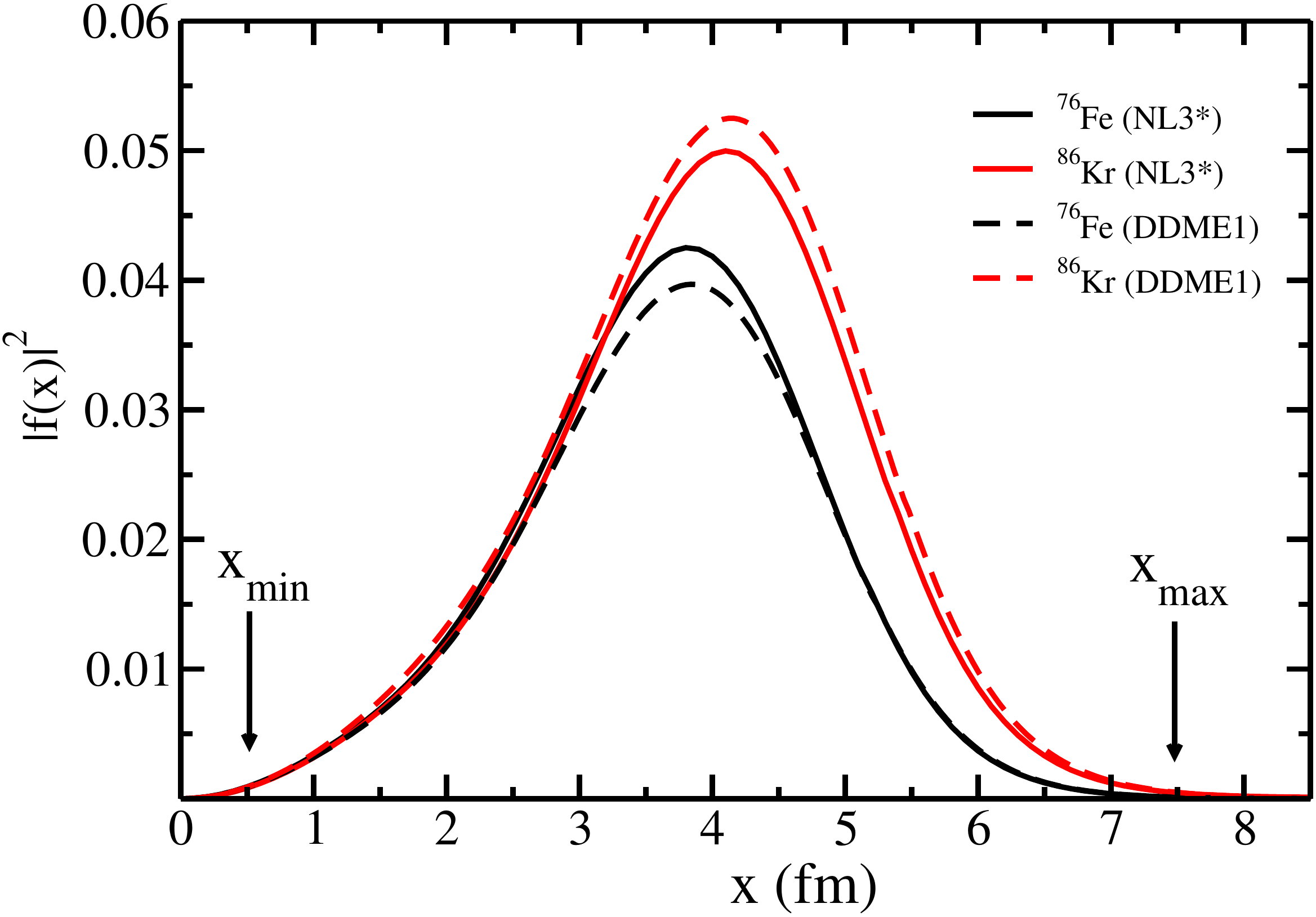}
\caption{\label{fig:2}(Color online) The weight function $\vert f (x) \vert^2$ calculated for $^{76}$Fe, and 
$^{86}$Kr isotopes by using the self-consistent NL3$^*$ and DD-ME1 total density distributions of these nuclei. 
The arrows stand for the minimum and maximum values of the integration limit taken in the subsequent 
calculations. See text for details.}
\end{center}
\end{figure}

The coherent density functional method (CDFM) was suggested and developed by Antonov {\it et al.} 
\cite{anto79,anto80}. It is based on the $\delta$-function limit of the generator coordinate method 
\cite{anto94,gai07,gai11}. In CDFM, the one-body density matrix $\rho$ ({\bf r}, {\bf r$'$}) of a 
finite nucleus can be written as a coherent superposition of the one-body density matrices $\rho_x$ 
({\bf r}, {\bf r$'$}) for spherical pieces of the nuclear matter called {\it fluctons},
\begin{equation}
\rho_x ({\bf r}) = \rho_0 (x) \Theta (x - \vert {\bf r} \vert), 
\label{denx} 
\end{equation}
with $\rho_o (x) = \frac{3A}{4 \pi x^3}$. The generator coordinate $x$ is the spherical radius of all 
$A$ nucleons contained in a uniform distributed spherical Fermi gas. In finite nuclear system, the one 
body density matrix is given as \cite{anto94,gai07,gai11,gai12}, 
\begin{equation}
\rho ({\bf r}, {\bf r'}) = \int_0^{\infty} dx \vert f(x) \vert^2 \rho_x ({\bf r}, {\bf r'}), 
\label{denr} 
\end{equation}
where, $\vert f(x) \vert^2 $ is the weight function (defined in Eq. (\ref{weight})). The term $\rho_x ({\bf r}, 
{\bf r'})$ is the coherent superposition of the one body density matrix and defined as,  
\begin{eqnarray}
\rho_x ({\bf r}, {\bf r'}) &=& 3 \rho_0 (x) \frac{J_1 \left( k_f (x) \vert {\bf r} - {\bf r'} \vert 
\right)}{\left( k_f (x) \vert {\bf r} - {\bf r'} \vert \right)} \nonumber \\
&&\times \Theta \left(x-\frac{ \vert {\bf r} + {\bf r'} \vert }{2} \right). 
\label{denrr}
\end{eqnarray}
Here, $J_1$ is the first order spherical Bessel function and $k_F (x)$ is the Fermi momentum of the nucleons 
in the flucton with radius $x$. The corresponding Wigner distribution function for the one body density 
matrices in Eq. (\ref{denrr}) is,
\begin{eqnarray}
W ({\bf r}, {\bf k}) =  \int_0^{\infty} dx \vert f(x) \vert^2 W_x ({\bf r}, {\bf k}), 
\label{wing}
\end{eqnarray}
where, $W_x ({\bf r}, {\bf k})=\frac{4}{8\pi^3}\Theta (x-\vert {\bf r} \vert)\Theta (k_F (x)-\vert {\bf k} 
\vert)$ . Similarly, the density $\rho$ (r) in the CDFM can express in terms of the same weight function 
as,
\begin{eqnarray}
\rho (r) &=& \int d{\bf k} W ({\bf r}, {\bf k}) \nonumber \\
&& = \int_0^{\infty} dx \vert f(x) \vert^2 \frac{3A}{4\pi x^3} \Theta(x-\vert {\bf r} \vert)
\label{rhor}
\end{eqnarray}
and it is normalized to the mass number, $\int \rho ({\bf r})d{\bf r} = A$. By taking the $\delta$-function 
approximation to the Hill-Wheeler integral equation, one obtains a differential equation for the weight 
function in the generator coordinate \cite{anto79,anto80,anto94}. We have adopted a conventional approach 
to the weight function instead of solving the differential equation (detail in Ref. \cite{anto80,anto94}). 
The weight function for a given density distribution $\rho$ (r) can be expressed as,
\begin{equation}
|f(x)|^2 = - \left (\frac{1}{\rho_0 (x)} \frac{d\rho (r)}{dr}\right )_{r=x}, 
\label{weight}
\end{equation}
with $\int_0^{\infty} dx \vert f(x) \vert^2 =1$. For a detailed analytical derivation, one can follow Refs. 
\cite{anto94,bro92,fuch95}. Here our principal goal is to define an effective symmetry energy, its slope, 
and curvature for a finite nucleus around by weighting the quantities for infinite nuclear matter within 
the CDFM. Following the CDFM approach, the effective symmetry energy $S_0$, its corresponding pressure $p_0$, 
and the curvature $K_0$ for a finite nucleus can be written as \cite{anto94,bro92,fuch95,gai07,gai11,gai12},
\begin{eqnarray}
S_0 = \int_0^{\infty} dx \vert f(x) \vert^2 S^{NM} (\rho (x)), \nonumber \\
p_0 =  \int_0^{\infty} dx \vert f(x) \vert^2 p_0^{NM} (\rho (x)), \nonumber \\
K_0 =  \int_0^{\infty} dx \vert f(x) \vert^2 K_0^{NM} (\rho (x)).  
\label{finite}
\end{eqnarray}
We will see that the quantities on the left-hand-side of Eq. (\ref{finite}) are surface weighted averages of 
the corresponding nuclear matter quantities $S^{NM}$, $p_0^{NM}$ and $K_0^{NM}$ on the right-hand-side. The 
region within $x_{min} \leq x \leq x_{max}$ (see Fig. \ref{fig:2} displaying the weigh function) is taken 
for the integration. More details can found in Section III. The calculated densities from the NL3$^*$ 
and the DD-ME1 are used for estimate the weight function $\vert f (x) \vert^2$ in Eq. (\ref{weight}) for 
each nucleus and used for the calculations in Eq. (\ref{finite}). The spin-independent proton and neutron 
mean-field densities are given by, 
\begin{eqnarray}
\rho ({\bf R}) = \rho (r_{\perp}, z) 
\label{rhoR}
\end{eqnarray}
where $r_{\perp}$ and $z$ are the cylindrical coordinates of the radial vector ${\bf R}$. The single 
particle densities are 
\begin{eqnarray}
\rho_i ({\bf R}) = \rho_i (r_{\perp}, z) = \vert \phi_i^+ (r_{\perp}, z) \vert^2 + 
\vert \phi_i^- (r_{\perp}, z) \vert^2, 
\label{spd}
\end{eqnarray}
where, $\phi_i^{\pm}$ is the wave function, expanded into the eigen functions of an axially symmetric deformed 
harmonic oscillator potential in cylindrical co-ordinates. The normalization of the densities is given by,
\begin{eqnarray}
\int \rho ({\bf R}) d{\bf R} = X,
\label{norm}
\end{eqnarray}
where X = N, Z for neutron and proton number, respectively. Further, the multipole decomposition of the 
density can be written in terms of even values of the multipole index $\lambda$ as,
\begin{eqnarray}
\rho (r_{\perp}, z) = \sum_{\lambda} \rho_i ({\bf R}) P_{\lambda} (Cos\theta).
\label{mdd}
\end{eqnarray}
Here, we have used the monopole term of the density distribution in the expansion Eq. (\ref{mdd}) for the 
calculation of the weight function $\vert f (x) \vert^2$ for simplicity. For a deformed nucleus, the peak 
of $\vert f (x) \vert^2$ does indeed depend on the angle. However, the density also depends on the angle 
in such a manner that the density at the peak of $\vert f (x) \vert^2$ is almost constant. The effect of 
the multipole component in the expansion can thus be neglected.  We can define the neutron skin thickness 
$\Delta R$ using the root-mean-square (rms) radii of neutrons and protons as,
\begin{eqnarray}
\Delta R = \langle r_n^2 \rangle - \langle r_p^2 \rangle. 
\label{skin}
\end{eqnarray}
The quantities defined above in Eq. (\ref{skin}) are used in the present study. 
\begin{table*}
\caption{The binding energy (BE), charge radius $r_{ch}$ and the quadrupole deformation parameter $\beta_2$ 
for the ground states of the $^{78-92}$Ge, $^{80-94}$Se and $^{82-96}$Kr nuclei for the non-linear NL3$^*$ 
and the density dependent DD-ME1 calculations compare with the experimental data \cite{audi12,ange13,prit12}, 
wherever available. The (*) marks in the binding energies of the experimental data are for extrapolated 
values.}
\renewcommand{\tabcolsep}{0.12cm}
\renewcommand{\arraystretch}{1.4}
\begin{tabular}{cccccccccc}
\hline\hline
Nucleus & \multicolumn{3}{c}{Binding Energy} & \multicolumn{3}{c}{Charge Radius}
& \multicolumn{3}{c}{Quadrupole Deformation} \\
& NL3$^*$ & DD-ME1 & Expt. \cite{audi12} & NL3$^*$ & DD-ME1 & Expt. \cite{ange13}
& NL3$^*$ & DD-ME1 & Expt. \cite{prit12} \\
\hline \hline
$^{76}$Ge &658.59&657.85&661.59     &4.052&4.050&4.0811 &0.171&0.179& $--$ \\
$^{78}$Ge &674.41&673.71&676.38     &4.064&4.061& $--$  &0.181&0.189&0.2623 \\
$^{80}$Ge &688.05&687.64&690.18     &4.071&4.066& $--$  &0.158&0.164& $--$ \\
$^{82}$Ge &699.53&699.56&702.43     &4.068&4.068& $--$  &0.001&0.012& $--$ \\
$^{84}$Ge &707.17&706.87&711.22$^*$ &4.099&4.101& $--$  &0.153&0.181& $--$ \\
$^{86}$Ge &714.42&714.24& $--$      &4.131&4.132& $--$  &0.207&0.216& $--$ \\
$^{88}$Ge &721.25&721.22& $--$      &4.160&4.161& $--$  &0.235&0.244& $--$ \\
$^{90}$Ge &726.52&726.92& $--$      &4.184&4.185& $--$  &0.235&0.236& $--$ \\
$^{92}$Ge &731.42&732.04& $--$      &4.206&4.208& $--$  &0.225&0.224& $--$ \\
\hline
\hline
$^{78}$Se &676.63&675.80&679.98     &4.113&4.110&4.1406 &0.162&0.181&0.2712 \\
$^{80}$Se &694.77&693.95&696.86     &4.122&4.119&4.1400 &0.173&0.185&0.2318 \\
$^{82}$Se &710.82&710.33&712.84     &4.128&4.126&4.1400 &0.154&0.170&0.1934 \\
$^{84}$Se &725.73&725.51&727.34     &4.123&4.117& $--$  &0.001&0.001& $--$ \\
$^{86}$Se &733.52&733.21&738.07     &4.141&4.151& $--$  &0.032&0.045& $--$ \\
$^{88}$Se &743.26&742.34&747.55     &4.185&4.187& $--$  &0.202&0.215& $--$ \\
$^{90}$Se &751.63&751.42&755.73$^*$ &4.216&4.214& $--$  &0.237&0.247& $--$ \\
$^{92}$Se &758.96&759.01&762.58$^*$ &4.241&4.240& $--$  &0.239&0.243& $--$ \\
$^{94}$Se &765.53&765.85&768.92$^*$ &4.266&4.266& $--$  &0.232&0.236& $--$ \\
\hline
\hline
$^{80}$Kr &691.74&691.08&695.43     &4.164&4.158&4.1970 &0.095&0.097&0.2650 \\
$^{82}$Kr &711.77&710.96&714.27     &4.171&4.167&4.1919 &0.124&0.126&0.2021 \\
$^{84}$Kr &730.13&729.56&732.25     &4.171&4.171&4.1884 &0.078&0.097&0.1489 \\
$^{86}$Kr &747.73&747.56&749.23     &4.174&4.170&4.1835 &0.001&0.001& $--$ \\
$^{88}$Kr &757.48&756.93&761.80     &4.192&4.195&4.2171 &0.027&0.105& $--$ \\
$^{90}$Kr &767.61&767.18&773.22     &4.227&4.228&4.2423 &0.158&0.173& $--$ \\
$^{92}$Kr &777.57&777.49&783.18     &4.261&4.269&4.2724 &0.210&0.236& $--$ \\
$^{94}$Kr &786.29&786.47&791.67$^*$ &4.286&4.289&4.3002 &0.218&0.222& $--$ \\
$^{96}$Kr &794.34&794.75&799.68$^*$ &4.307&4.311&4.3267 &0.206&0.208& $--$ \\
\hline \hline
\end{tabular}
\label{Tab3}
\end{table*}
\begin{figure}
\begin{center}
\includegraphics[width=1.00\columnwidth]{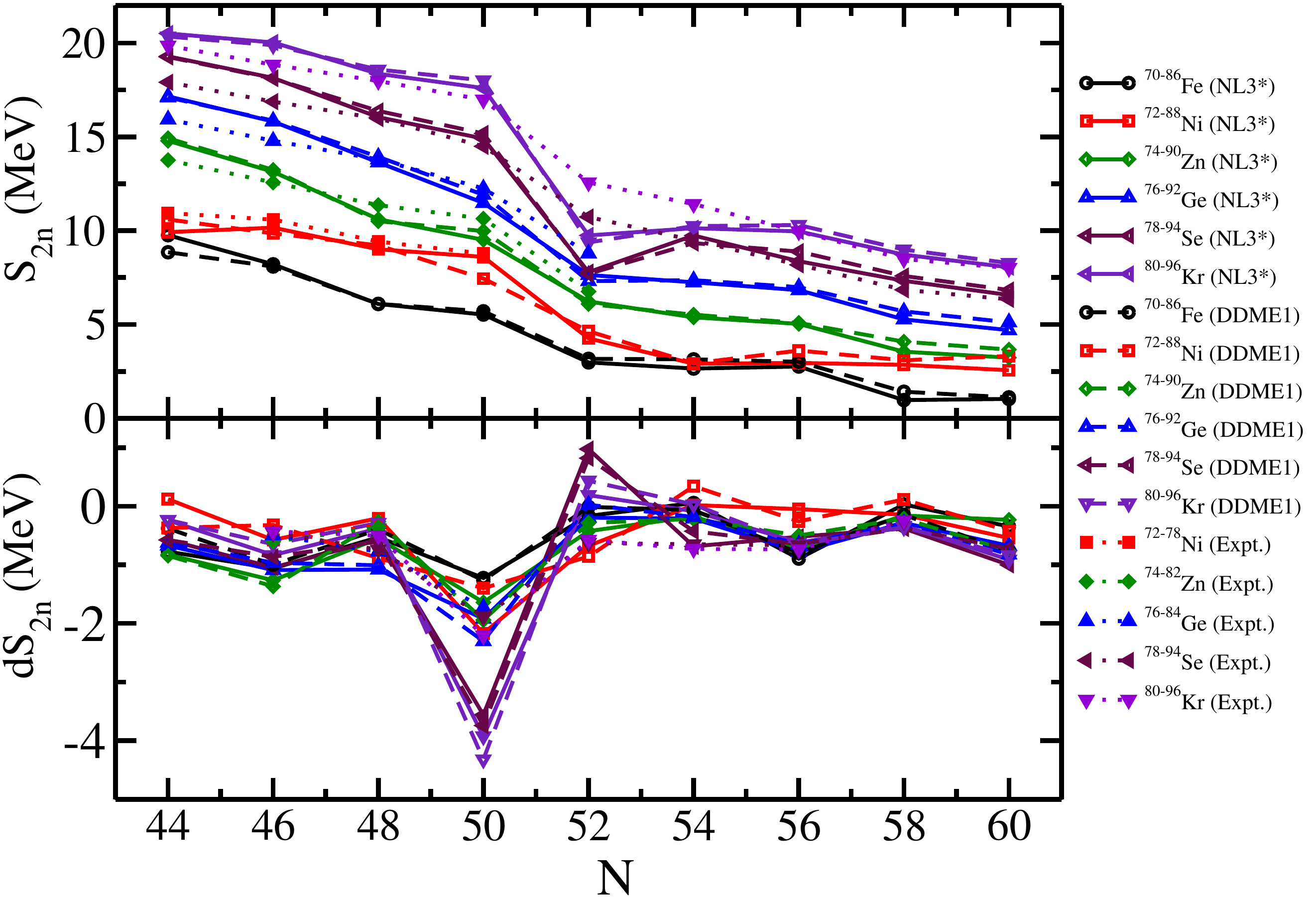}
\caption{\label{fig:01} (Color online) The two neutron separation energies $S_{2n}$ (upper panel) and 
the differential variation of the separation energy $dS_{2n}$ (lower panel) from the NL3$^*$ and the 
DD-ME1 interactions are given for Fe, Ni, Zn, Ge, Se, and Kr isotopic chains. The experimental datas 
\cite{audi12} are given for comparison, wherever available. See text for details.}
\end{center}
\end{figure}

\section{Calculations and Results}
In the relativistic mean field model, the field equations are solved self-consistently by taking different 
inputs for the initial deformation $\beta_0$ \cite{lala97,lala09,ring86,ring90,bhu09,niks02,bret00}. To 
verify the convergence of the ground state solutions for this mass region, we performed calculation for the 
number of major boson shells $N_B$ =16 and varied the number of major fermion shells $N_F$ from 10 to 20. 
From the results obtained, we have confirmed that the relative variations of these solutions are $\leq$ 
0.004$\%$ for the binding energy and 0.001$\%$ for the nuclear radii over the range of major fermion shells. 
Hence, the desired number of major shells for  fermions and bosons were fixed at $N_F$ = 16 and $N_B$ = 16. 
The number of mesh points for Gauss-Hermite and Gauss-Laguerre integration used are $20$ and $24$, 
respectively. For a given nucleus, the solution corresponding to the maximum binding energy is treated as 
the ground state and other solutions are considered excited states of the nucleus. We have used the 
non-linear NL3$^*$ \cite{lala09} and density-dependent DD-ME1 \cite{niks02} interactions for the present 
analysis. These interaction parameters are widely used and are able to provide a reasonable good description 
of the properties of nuclei from light to super-heavy, from the proton to the neutron drip line 
\cite{lala09,bhu09,bhu11,bhu15}. The calculations furnish the ground state bulk properties such as binding 
energy, rms charge radius, nuclear qudrupole deformation $\beta_2$, nuclear density distribution 
$\rho (r_{\perp}, z)$, and the single particle energy. 

The results obtained from both sets of interaction parameters along with the experimental data 
\cite{audi12,ange13,prit12} are listed in Tables. \ref{Tab2} and \ref{Tab3}. From the tables, one notices 
that the results of our calculations agree quite well with the experimental data for binding energy 
and root-mean-square charge radius, wherever available. In both the NL3$^*$ and DD-ME1 results, we find 
deformed prolates solutions in the ground state configuration for entire isotopic chains except the mass 
region near N = 50. In other words, the deformed prolate configuration follows a spherical solution at 
N = 50 and again becomes deformed with increasing neutron number for the Fe, Zn, Ge, Se and Kr nuclei. In 
the case of the isotopic chain of Ni, we found almost spherical solutions for the entire isotopic chain, 
which do not appear in case of other nuclei (see the Tables \ref{Tab2} and \ref{Tab3}). The experimental 
data for the deformations are slightly underestimated by the calculations for both sets of interaction 
parameters. 

\begin{figure}
\begin{center}
\includegraphics[width=1.00\columnwidth]{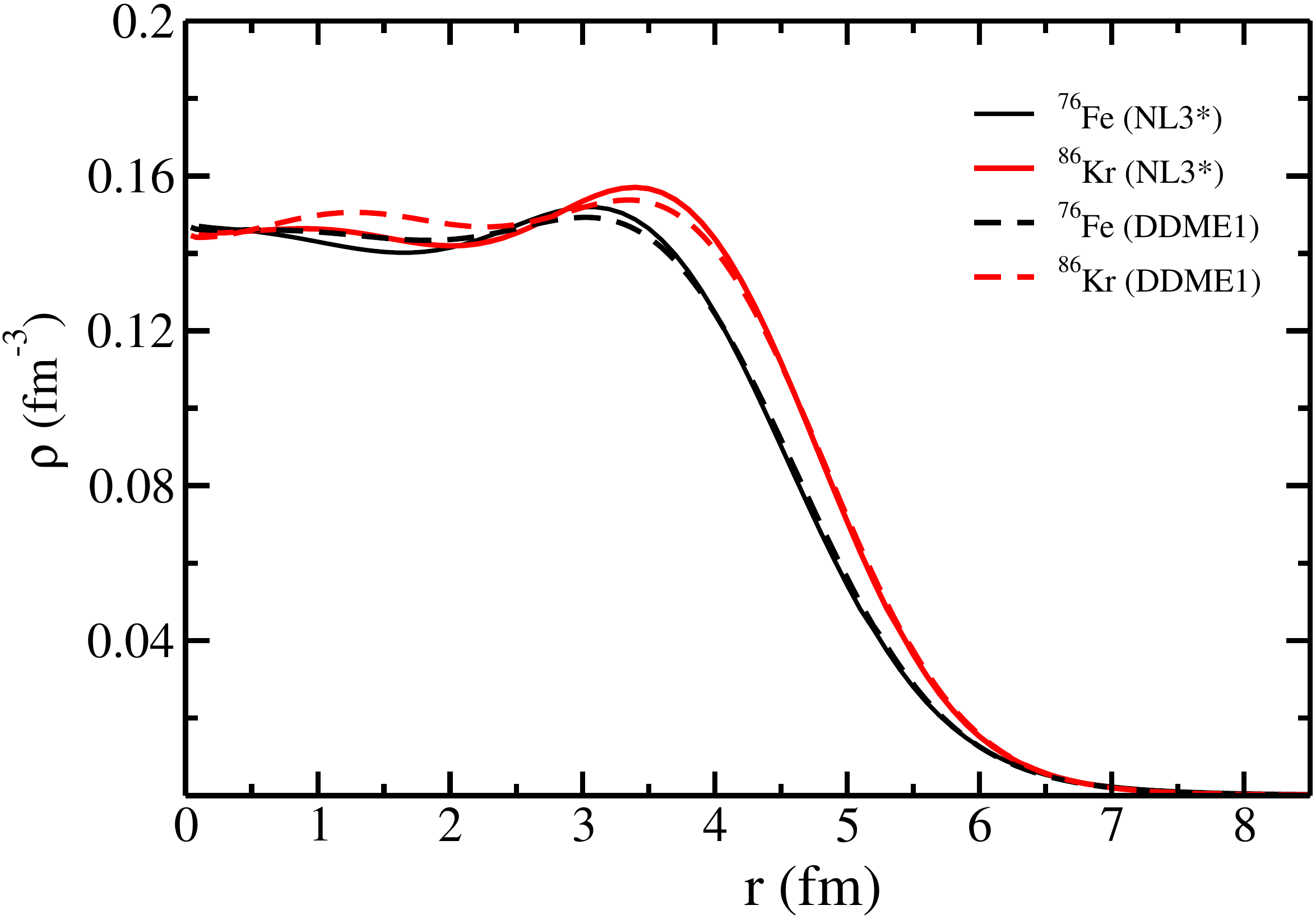}
\caption{\label{fig:1} (Color online) The microscopic relativistic mean field with non-linear NL3$^*$ and 
Dirac-Hartree-Bogoliubov with DD-ME1 total density distribution for $^{76}$Fe, and ${86}$Kr isotopes. See 
text for details.}
\end{center}
\end{figure}

\subsection{The neutron separation energy}
The two neutron separation energy  $S_{2n}$ (Z, N), can be estimated from the ground state nuclear masses 
$BE$ (Z, N) and $BE$ (Z, N-2) and the neutron mass $m_n$ by the relation:
\begin{eqnarray}
S_{2n} (Z, N) = -BE (Z, N) + BE (Z, N-2) + 2m_n,
\end{eqnarray}
The $BE$ of the $^AZ$ and $^{A-2}Z$ nuclei are calculated using the non-linear NL3$^*$ and the 
density-dependent DD-ME1 interaction parameters. Here, we have used the experimental datas \cite{audi12} 
to obtain the experimental values of the $S_{2n}$ energy for comparison with our calculated results. From 
the Tables \ref{Tab2} and \ref{Tab3}, one observes that the calculated binding energies are reasonably good 
agreement with the available experimental data, which shows the predictive power of the calculations for the 
correct evolution of shell structures. The estimated $S_{2n}$ results along with the experimental values 
(wherever available) are shown in the upper panel of Fig. \ref{fig:01} for $^{70-86}$Fe, $^{72-88}$Ni, 
$^{74-90}$Zn, $^{76-92}$Ge, $^{78-94}$Se, and $^{80-96}$Kr isotopes. The two-neutron separation energy 
$S_{2n}$, as a function of the neutron number in an isotopic chain, decreases smoothly as the number of 
neutron increases. Sharp discontinuities (kinks) appear at the neutron spherical closure magic number 
$N$ = 50. In terms of energy, the energy necessary to remove two neutrons from a nucleus (Z, $N_{magic}$+2) 
is much smaller than that to remove two neutrons from the nucleus (Z, $N_{magic}$), which breaks the regular 
trend. From Fig. \ref{fig:01} (upper panel), one can observe that the calculated results are in agree with 
the experimental data and also follow the expected trend along the isotopic chains. 

To better explore the dependence of $S_{2n}$ with respect to neutron number, the differential variation 
of the $S_{2n}$ ($dS_{2n}$ (N,Z)) is defined as
\begin{eqnarray}
dS_{2n}(Z,N)=\frac{S_{2n}(Z,N+2)-S_{2n}(Z,N)}{2},
\end{eqnarray}
In the Fig. \ref{fig:01} (upper panel), we observe that the curves for isotopic chains for different atomic 
number shows roughly the similar trends. From these general characteristics of the $S_{2n}$ curves we expect 
that the derivative, $dS_{2n}$, should have a sharp fall in the negative direction for magic or/and 
semi-magic neutron number in an isotopic chain. In other words, the magnitude of the sharp drop, at magic 
neutron numbers shows the strength of the shell structure for that specific neutron number in the isotopic 
chain. Here, we found similar characteristics for the Fe, Ni, Zn, Ge, Se and Kr nuclei (see the lower panel 
of Fig. \ref{fig:01}). The experimental values \cite{audi12} are also given for comparison. Further, the 
depth of $dS_{2n}$ at magic neutron number increases along the isotonic chain. The fall in $dS_{2n}$ at $N$ 
= 50 for the isotopic chain discloses additional nuclear structure features. 

\subsection{The Nuclear Density and Weight Function}
Once we have the density in hand, we estimate the nuclear matter observables using these densities in 
the framework of the coherent density functional method (CDFM) \cite{anto79,anto80,gai07,gai11,gai12}, 
which involves the following steps: (i) we generate the weight function $\vert f (x) \vert^2$ for each 
nucleus using the density distribution obtained from the RMF (NL3$^*$ and DD-ME1), as defined in Eq. 
(\ref{weight}) \cite{anto79,anto80,gai07,gai11,gai12}, (ii) then we use this weight function along with 
the nuclear matter observables to calculate the effective symmetry energy properties in finite nuclei 
\cite{anto79,anto80,gai07,gai11,gai12}. We compare our calculated results with other theoretical predictions
and examine the influences of these observables on the prediction of shell closures in each isotopic chain 
and the constraints they place on nuclear matter observables. In Fig. \ref{fig:1}, we have plotted the total 
density distribution (sum of the proton $\rho_p$ and the neutron $\rho_n$ density) for $^{76}$Fe, and 
$^{86}$Kr obtained from the NL3$^*$ and DD-ME1 interaction parameters as a representative case. One finds 
similar characteristics of the density for all nuclei considered in the present study. Further, a careful 
inspection shows a small enhancement in the surface region with an increase in proton number. In other words, 
the total density is extended towards the tail region in an isotonic chain and this ostensible distinction 
plays a significant role in the effective nuclear matter quantities.  

The weight functions (in Eq. (\ref{weight})) is interlinked with the nuclear matter observables, such as the 
symmetry energy, the neutron pressure and their related observables \cite{anto79,anto80,gai07,gai11,gai12}. 
Following the CDFM approach, we briefly discuss the weight function $\vert f (x) \vert^2$ [i.e. in Eq. 
(\ref{weight})], which is directly associated with the density distribution of the finite nucleus. We have 
estimated the weight function of each nucleus using its total density ($\rho_p + \rho_n$) distribution 
obtained from the relativistic mean field model. Here, we have given the $\vert f (x) \vert^2$ for $^{76}$Fe, 
and $^{86}$Kr nuclei as representative cases, which are shown in Fig. \ref{fig:2}. The weight function is the 
crucial quantity for describing the surface properties of the finite nucleus in terms of effective nuclear 
matter quantities. One can see from the figure, the weight function has a peak near the surface of the 
nuclear density density distribution. In other words, one finds a peak in the weight function $\approx$ 5 
$fm$, which is due to contributions from the surface region of the nuclear density. For a better 
comprehension of this fact, one should compare the plots of the density distribution to those of the weight 
factor (i.e. see Figs. \ref{fig:1} and \ref{fig:2}).

As we mentioned above, the objective of the present investigation is to study correlations between the 
neutron-skin thickness and effective nuclear matter properties such as the symmetry energy, neutron pressure 
(proportional to the slope of the bulk symmetry energy), and curvature in a given isotopic chain. Following 
Eq. (\ref{finite}), we first introduce the value of $x_{min}$ at which the symmetry energy for nuclear 
matter $S^{NM}$ (x) changes sign from negative to positive at $x_{min} \geq x \leq x_{max}$ (see Fig. 
\ref{fig:2}). In other words, the $S^{NM} < 0$ for the values of $ x \leq x_{min}$ and $ x \geq x_{max}$ 
in Eq. \ref{finite}. Considering the basic principle of the CDFM, the domain of $x$ should run from $0$ 
to $\infty$, which incorporates the region of densities $\rho_0$ (x) from $\infty$ to $0$, as well. At a 
point where the value of $x$ is very small, in practice the estimate provides the values of density $\rho_0$ 
(x) that are much larger than the saturation density. To avoid such a nonphysical situation (i.e. a negative 
value of the symmetry energy), we include the value of $x \geq x_{min}$ for the lower limit and 
simultaneously exclude $x \geq x_{max}$ from the upper limit of the integration in Eq. (\ref{finite}). The 
estimated values of $x_{min}$ and $x_{max}$ of the integration are shown in Fig. \ref{fig:2}. 

\begin{figure}
\begin{center}
\includegraphics[width=1.00\columnwidth]{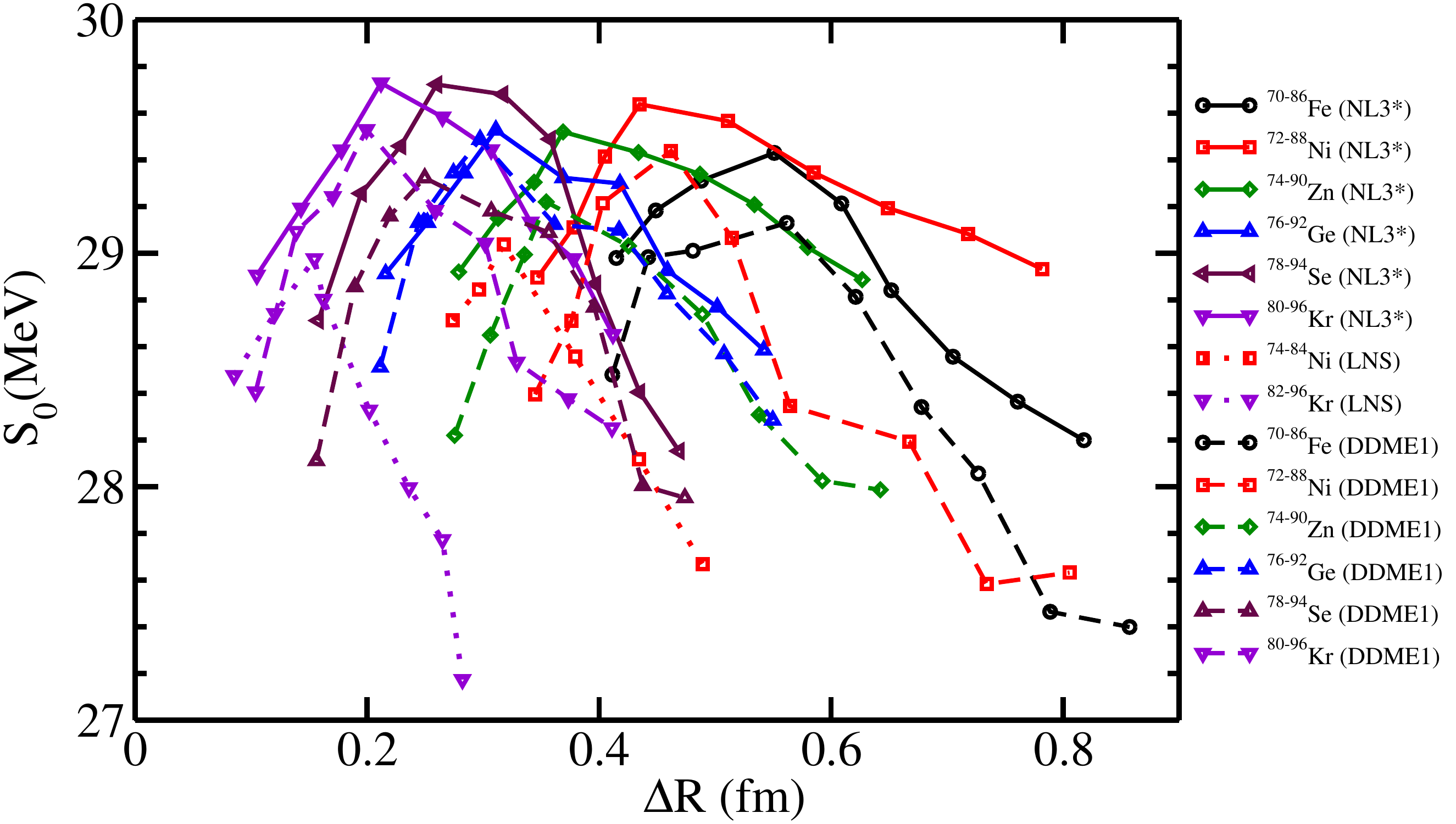}
\caption{\label{fig:3} (Color online) The symmetry energy $S_0$ for $^{70-86}$Fe, $^{72-88}$Ni, $^{74-90}$Zn, 
$^{76-92}$Ge, $^{78-94}$Se, and $^{80-96}$Kr isotopes as a function of the neutron skin thickness $\Delta R$ 
as calculated using the RMF NL3$^*$ (solid line) and DD-ME1 (dashed line) interactions. The Skyrme-Hartree-Fock 
+ BCS results for the LNS interaction \cite{gai11,gai12} (dotted line) are given for comparison, where 
available. See the text for details.}
\end{center}
\end{figure}
\begin{figure}
\begin{center}
\includegraphics[width=1.00\columnwidth]{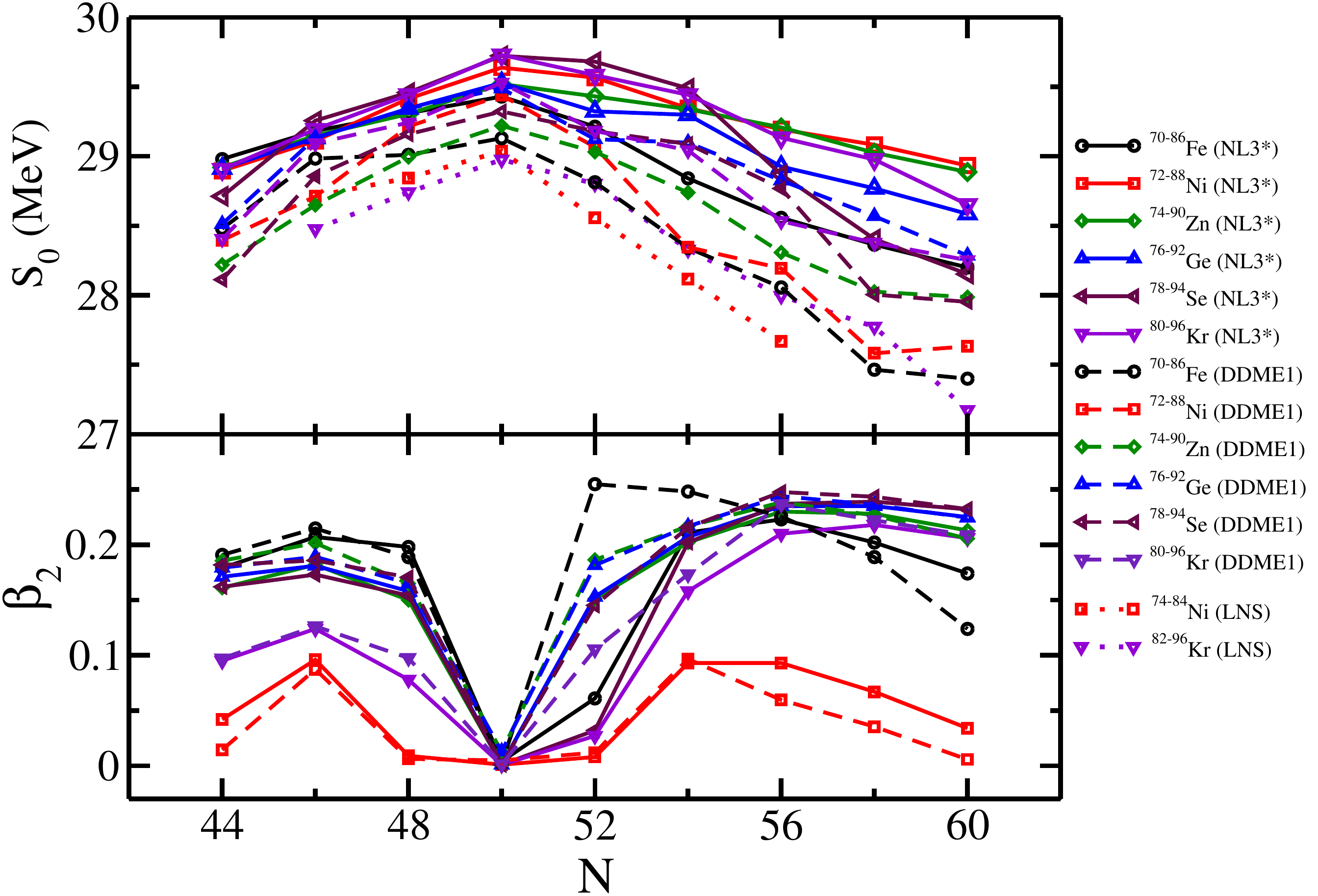}
\caption{\label{fig:4} (Color online) The symmetry energy $S_0$ and the quadrupole deformation $\beta_2$ for 
$^{70-86}$Fe, $^{72-88}$Ni, $^{74-90}$Zn, $^{76-92}$Ge, $^{78-94}$Se, and $^{80-96}$Kr isotopes as a function 
of neutron number $N$ are given in the upper and lower panel, respectively. The calculated results from RMF 
NL3$^*$ (solid line) and DD-ME1 (dashed line) interactions are compared with the Skyrme-Hartree-Fock + BCS 
results for the LNS interaction \cite{gai11,gai12} (dotted line) where available. See the text for details.}
\end{center}
\end{figure} 

\subsection{The Symmetry Energy}
The RMF calculations furnish principally nuclear structure properties, such as the quadrupole moment 
$Q_{20}$, nucleon density distribution $\rho (r_{\perp},z) = \rho_p (r_{\perp},z) +\rho_n(r_{\perp},z)$, 
and the root-mean-square nuclear radii. We estimate the neutron skin thickness $\Delta R$ of nuclei in 
a given isotopic chain using the neutron and proton radii obtained from the relativistic mean field with 
the non-linear NL3$^*$ and density-dependent DD-ME1 interaction parameters. The symmetry energy $S_0$ 
for a given nucleus is calculated within the CDFM through the weight function $\vert f (x) \vert^2$ (obtained 
from the self consistent density distribution) using Eq. (\ref{finite}). We show the symmetry energy $S_0$ 
as a function of neutron skin thickness in Fig. \ref{fig:3} for the $^{70-86}$Fe, $^{72-88}$Ni, $^{74-90}$Zn, 
$^{76-92}$Ge, $^{78-94}$Se, and $^{80-96}$Kr nuclei using the NL3$^*$ (solid line) and DD-ME1 (dashed line) 
interactions. The results obtained from a Skyrme-Hartree-Fock (SHF) + BCS with LNS interaction are also 
given for comparison, where available. From the figure, we observe a smooth growth of $S_0$ up to the neutron 
number (N = 50) and then a linear decrease of $S_0$, where the neutron-skin thickness of the isotopes 
increases. The SHF displays a similar behavior of the symmetry energy with respect to the skin thickness. 
Careful inspection shows that the neutron skin thicknesses obtained from the RMF (NL3$^*$ and DD-ME1) are 
slightly larger when compared to those of the LNS interaction parameter. Further, the values of the $S_0$ 
for the relativistic interactions are slightly larger than the non-relativistic LNS predictions, which can 
reflect on the nuclear matter characteristics \cite{dutra12,dutra14}. 

The results exhibited in Fig. \ref{fig:3} are shown from an additional point of view in Fig. \ref{fig:4}. 
In the upper and the lower panels of Fig. \ref{fig:4}, we give the evolution of the symmetry energy and 
the quadrupole deformation $\beta_0$ as a function of the mass number, respectively. From the figure, we 
observe a similar peak of the symmetry energy at N = 50 for all the isotopic chains (see the upper panel 
of Fig. \ref{fig:4}). One sees in Figs. \ref{fig:3} and \ref{fig:4} that $S_0$ varies by about 29.0 $\pm$ 
1.0 MeV in the interval for the NL3$^*$ and DD-ME1 interaction parameters. The LNS interaction yields a 
values of $S_0$ smaller by $\approx 1$ unit than the corresponding values of the relativistic interactions. 
The evolution of the symmetry energy is related to the development of the quadrupole moment as a function 
of the mass number, as displayed in the lower panel of Fig \ref{fig:4}. From the trajectory of the 
quadrupole deformation parameter $\beta_2$ as a function of mass number, one can see that the semi-magic 
isotopes corresponding to the neutron number N = 50 are spherical for both the NL3$^*$ and DD-ME1 
interactions, while the open-shell isotopes within these isotopic chains have a prolate ground state 
configuration. Following Fig. \ref{fig:4}, one can clearly see that the peak of the symmetry energy 
occurs for the closed shell nuclei that are spherical in shape. The open-shell nuclei display a slight 
decrease of the symmetry energy along the deformed shell. This represent a possible direction for further 
systematic investigation of the isospin dependence of the nuclear equation of state.

\begin{figure}
\begin{center}
\includegraphics[width=1.00\columnwidth]{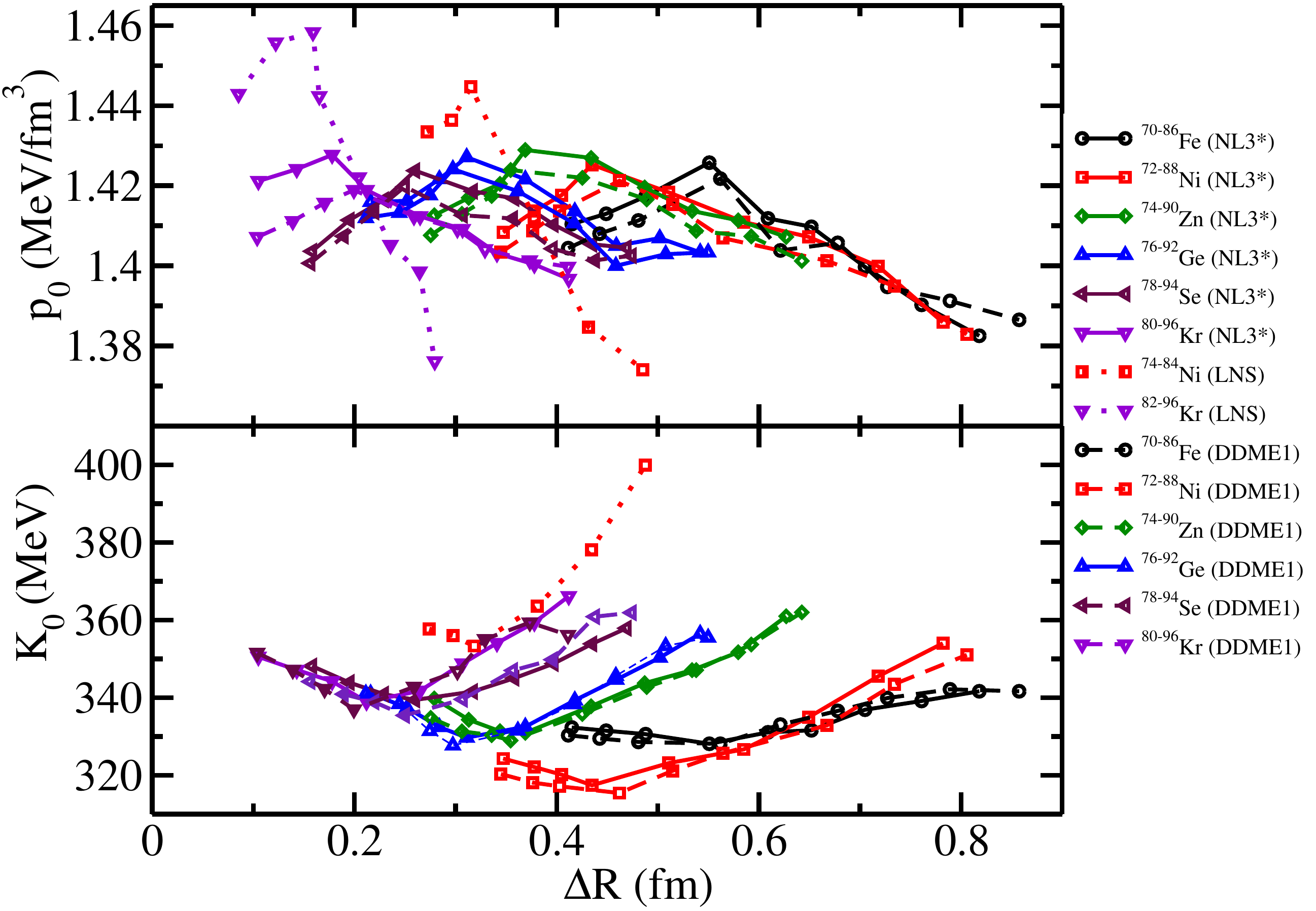}
\caption{\label{fig:5}(Color online) The neutron pressure $p_0$ and the curvature $K_0$ for $^{70-86}$Fe, 
$^{72-88}$Ni, $^{74-90}$Zn, $^{76-92}$Ge, $^{78-94}$Se, and $^{80-96}$Kr isotopes as a function of neutron 
skin thickness $\Delta R$ using the non-linear NL3$^*$ (solid line) and density-dependent DD-ME1 (dashed 
line) interactions are displayed in the upper and lower panels, respectively. The Skyrme-Hartree-Fock 
results for the LNS interaction \cite{gai11,gai12} (dotted line) are given for comparison, where available. 
See text for details.}
\end{center}
\end{figure}
\begin{figure}
\begin{center}
\includegraphics[width=1.00\columnwidth]{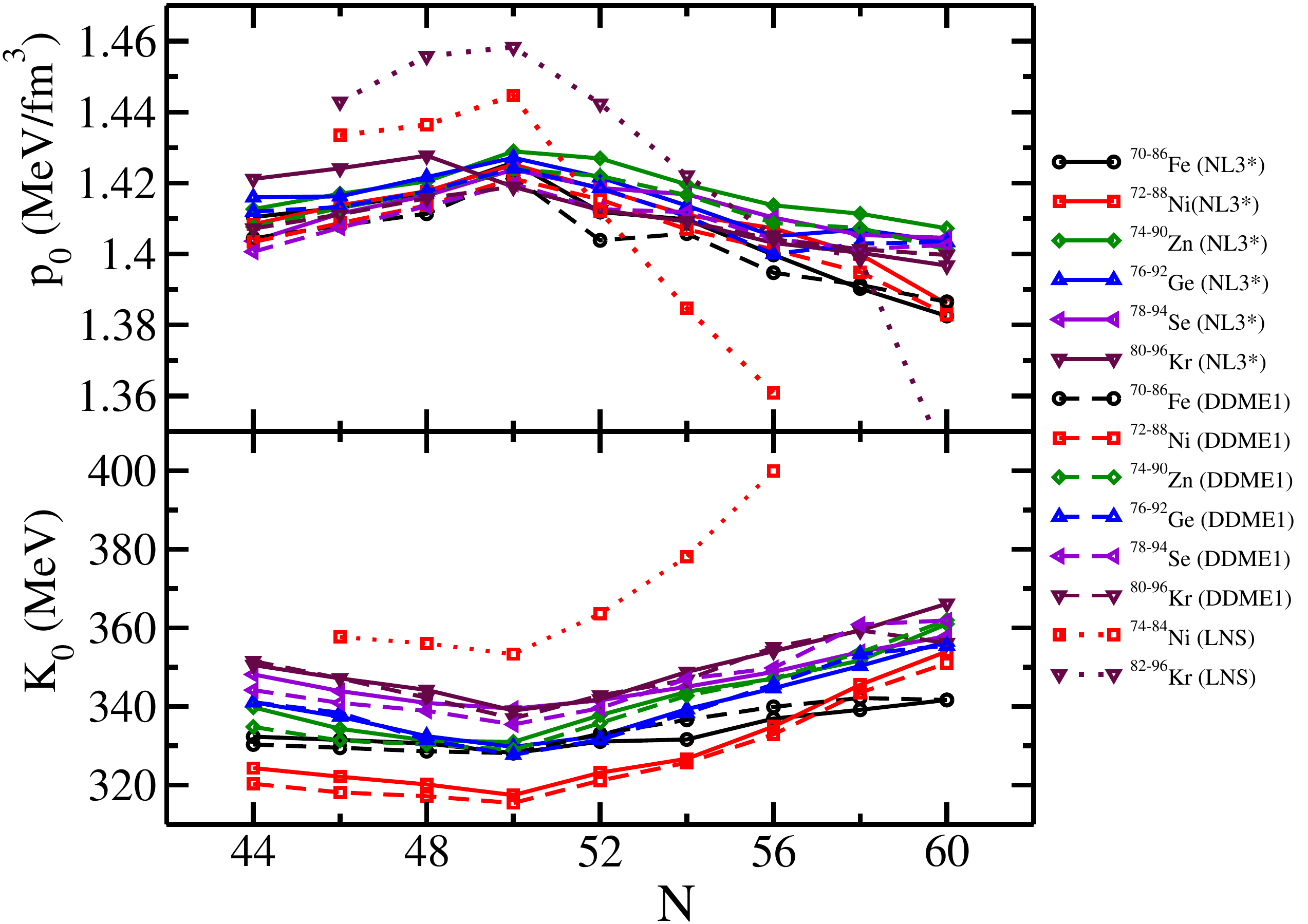}
\caption{\label{fig:6}(Color online) The neutron pressure $P_0$ and the curvature $K_0$ for $^{70-86}$Fe, 
$^{72-88}$Ni, $^{74-90}$Zn, $^{76-92}$Ge, $^{78-94}$Se, and $^{80-96}$Kr isotopes as a function of mass 
number for the non-linear NL3$^*$ and density-dependent DD-ME1 interactions are given in the upper and 
lower panels, respectively. The Skyrme-Hartree-Fock results for the LNS interaction \cite{gai11,gai12} 
(dotted line) are also given for comparison, where available. See the text for details.}
\end{center}
\end{figure}

\subsection{Neutron Pressure and Curvature}
Next, we illustrate a possible correlation of the neutron skin thickness $\Delta$R with the neutron 
pressure $p_0$ and the curvature $K_0$, in Fig. \ref{fig:5}. In Fig \ref{fig:6}, we plot the trajectory 
of $p_0$ and $K_0$ with neutron number for the Fe, Ni, Zn, Ge, Se and Kr nuclei. The calculated results 
from the RMF NL3$^*$ (solid line) and DD-ME1 (dashed line) interactions are given with the results of 
the non-relativistic Skyrme-Hartree-Fock (LNS) interaction (dotted line) \cite{gai11,gai12} for comparison, 
where available. From the figure it is clearly seen that the neutron skin thickness of the isotopes 
correlates almost linearly below and above the minimum with $p_0$ and $K_0$, as does $S_0$. Similar to the 
symmetry energy, here we also find a peak in the neutron pressure $p_0$ and a minimum in the curvature 
$K_0$ for semi-magic nuclei at N = 50 for both interactions. Further, a slightly distorted transition is 
found in the linear correlation as compared to that of the symmetry energy curve. The small difference in 
the linear behavior indicates that the stability pattern is not as regular for the isotopic chain. As we 
have mentioned above, the peak follows a valley for a transition from a closed shell to an open shell 
nuclei. Here we have also found the same variation in the neutron pressure and curvature in the isotopic 
chains. It is worth mentioning that the decrease in $S_0$, $p_0$ and $K_0$ in the case of open-shell nuclei 
is due to the different occupancies of the single particle levels. Hence, we see that in general peaks are 
produced at shell closures. However analysis of the precise dependence of the various peaks on the occupation 
number of specific shells will require further work. The results obtained from the non-linear NL3$^*$ and 
density-dependent DD-ME1 interactions for $p_0$ and  $K_0$ show a similar trend to that of the LNS force. 
More careful inspection shows that the results for $p_0$ and  $K_0$ from our calculations are slightly smaller 
values than those of the LNS predictions. As we know, the magicity and/or shell closure (s) in an isotopic 
and/or isotonic chain are universal properties as far as the model used. Here, we get similar trends for 
non-linear NL3$^*$ and density-dependent DD-ME1 interactions, which also qualitatively agree with the 
non-relativistic NLS predictions. Hence, we can conclude, the results obtained in the present calculations 
are fairly model independent.  

\section{Summary and Conclusions}
In the present study, we have investigated possible relationships between the neutron skin thickness of 
neutron-rich nuclei and nuclear matter characteristics. A microscopic approach based on an axial deformed 
relativistic mean field with the non-linear NL3$^*$ and density-dependent DD-ME1 interaction parameters 
has been used. Effective nuclear matter properties such as the symmetry energy $S_0$, the neutron pressure 
$p_0$ and the nuclear curvature parameter $K_0$ have been determined for finite nuclei. The coherent density 
functional method was used to provide a transparent and analytic manner of calculating the effective 
infinite nuclear matter quantities by means of a weight function. In the first step, we have obtained the 
ground state nuclear bulk properties such as the binding energies, quadrupole deformations, nuclear density 
distributions using the self-consistent microscopic RMF with the NL3$^*$ and DD-ME1 interactions. We have 
considered the even$-$even isotopic chains of Fe, Ni, Zn, Ge, Se and Kr nuclei in the present analysis. The 
two neutron separation energies and the differential variation of the separation energies are also estimated 
from the microscopic binding energy for both the sets interaction parameters. From the separation energies, 
we found shell closures at N =50 for all the isotopic chains considered for both interactions. The neutron 
skin thickness and the weight function for each nucleus were estimated using the root-mean-square radius and 
the total density distribution, respectively. 

In the second step, we have calculated effective infinite nuclear matter characteristics such as symmetry 
energy $S_0$, neutron pressure $p_0$ and curvature $K_0$ for the finite nuclei. For all of the isotopic 
chains, we found that there exists a strong correlation between the neutron skin thickness and the symmetry 
energy. We found a peak in $S_0$ in an isotopic chain, which corresponds to the semi-magic isotopes at 
N = 50 and a spherical solution. An inflection-point transition appears for deformed nuclei at the spherical 
shell closures for the semi-magic isotopes at N =50 in the isotopic chain. In addition to these, a similar 
correlation between $\Delta R$ versus $p_0$ and $\Delta R$ versus $K_0$ has also been observed in the 
isotopic chains for both the NL3$^*$ and DD-ME1 sets of interaction parameters. The effect of the relative 
neutron-proton asymmetry on the evolution of the symmetry energy has been pointed out for these isotopes in 
the range $44 \leq N \leq 60$. We observe that the microscopic theoretical approach used is capable of 
predicting additional nuclear matter quantities in neutron-rich exotic nuclei and their connection to the 
surface properties of these nuclei. New exploratory results on giant resonances and the neutron skin in 
heavy nuclei and heavy-ion collisions could lead to new constraints on the nuclear symmetry energy, 
permitting an increased understanding of the physical quantities of nuclear systems.

\section*{Acknowledgments}
This work has been supported by FAPESP Project Nos. (2014/26195-5 \& 2017/05660-0), INCT-FNA Project No. 
464898/2014-5, the 973 Program of China (Grant No. 2013CB834400), the Chinese Academy of Sciences (Grant No. 
KJCX2-EW-N01), and by the CNPq - Brasil.

\end{document}